\documentclass[10pt,journal,final]{IEEEtran}

\ifCLASSOPTIONcompsoc
   \usepackage[nocompress]{cite}
\else
   \usepackage{cite}
\fi
\usepackage{graphicx}
\usepackage{epstopdf}
\usepackage{psfrag}
\usepackage{lscape}
\usepackage{amssymb}
\usepackage{amsmath}
\usepackage{cases}
\usepackage{ctable}
\usepackage{verbatim}
\usepackage{mathrsfs}
\usepackage{color,soul}
\ifCLASSOPTIONcompsoc
\usepackage[tight,normalsize,sf,SF]{subfigure}
\else
\usepackage[tight,footnotesize]{subfigure}
\fi

\begin{document}
%
\title{Least Squares Estimation-Based Synchronous Generator Parameter Estimation Using PMU Data}
\author{Bander Mogharbel, Lingling Fan, ~\IEEEmembership{Senior Member, IEEE}, Zhixin Miao, ~\IEEEmembership{Senior Member, IEEE} 
\thanks{B. Mogharbel, L. Fan and Z. Miao are with Smart Grid Power Systems Lab at EE Dept of University of South Florida. Email: linglingfan@usf.edu. }}
\maketitle

\begin{abstract}
In this paper, least square estimation (LSE)-based dynamic generator model parameter identification is investigated. Electromechanical dynamics related parameters such as inertia constant and primary frequency control droop for a synchronous generator are estimated using Phasor Measurement Unit (PMU) data obtained at the generator terminal bus. The key idea of applying LSE for dynamic parameter estimation is to have a discrete \underline{a}uto\underline{r}egression with e\underline{x}ogenous input (ARX) model. With an ARX model, a linear estimation problem can be formulated and the parameters of the ARX model can be found. This paper gives the detailed derivation of converting a generator model with primary frequency control into an ARX model. The generator parameters will be recovered from the estimated ARX model parameters afterwards. Two types of conversion methods are presented: zero-order hold (ZOH) method and Tustin method. Numerical results are presented to illustrate the proposed LSE application in dynamic system parameter identification using PMU data.
\end{abstract}




\section{introduction}
\label{intro}

Traditionally, Supervisory Control and Data Acquisition (SCADA) system using nonsynchronous data with low density sampling rate is used for monitoring and control of the system. Such measurements can not capture the system dynamics. The advent of phasor measurement units (PMUs) equipped with GPS antenna provides voltage/current phasors and frequency with a high density sampling rate up to 60 Hz. These phasor measurements transmitted with time stamps can help control systems have an accurate picture of the power system.\par

Synchronous generator parameter estimation has been investigated in the literature. Based on the scope of estimation, some only investigate electrical state estimation (e.g. rotor angle and rotor speed) \cite{yang2013power, Huang2009Dynamic}, while others estimate both system states and generator parameters \cite{Huang2009Application, Haung2006Model1, Huang2007Feasibility,Huang2011calibrating }. Based on estimation methods, there are at least two major systematic methods for parameter estimation: least squares estimation (LSE) \cite{ref6, ref8, ref9} and Kalman filter-based estimation \cite{Huang2013Generator,fan2013extended, Kamwa2011Dynamic, Kamwa2011Online, Pal2014pwrs_UKF}. To use LSE for dynamic system parameter estimation, a window of data is required. On the other hand, Kalman filter-based estimation can carry out estimation procedures at each time step. Thus Kalman filter-based estimation can be used for online estimation.

Majority of the research papers on PMU-based dynamic parameter estimation adopt Kalman filter-based estimation approach,\emph{ e.g}., \cite{Huang2013Generator,fan2013extended, Kamwa2011Dynamic, Kamwa2011Online,Pal2014pwrs_UKF }. In the literature, there are plenty of research on LSE-based offline generator parameter estimation based on either time-domain data or frequency response data \cite{ref6, ref7, ref8, ref9, ref12, ref14, ref15, ref16}. However, little research can be found for LSE-based dynamic parameter estimation for PMU data-based applications. In addition, existing research on LSE-based generator parameter estimation uses nonlinear LSE \cite{ref6}, existing toolboxes for nonlinear system identification (e.g., Matlab System Identification Toolbox) \cite{ref8}. The adoption of Matlab toolbox does not provide the insights as how the algorithms work, while the nonlinear LSE often time faces convergence issues if the set of the parameters chosen are not suitably chosen as demonstrated in \cite{ref6}.

Therefore, this paper aims to adopt linear LSE for generator parameter estimation via PMU data. Electromechanical dynamics related parameters such as inertia constant and primary frequency control droop for a synchronous generator are estimated using PMU data obtained at the generator terminal bus. The key idea of applying LSE for dynamic parameter estimation is to have a discrete ARX model. With an ARX model, a linear estimation problem can be formulated and the parameters of the ARX model can be found. This paper gives the detailed derivation of converting a generator model with primary frequency control into an ARX model. The generator parameters will be recovered from the estimated ARX model parameters afterwards. Two types of conversion methods are presented: zero-order hold (ZOH) method and Tustin method. Numerical results are presented to illustrate the proposed LSE application in dynamic system parameter identification using PMU data.

The rest of the paper is organized as follows. Section II describes the LSE for an ARX model. Section III describes the two approaches to convert a continuous transfer function into a discrete ARX model. Section IV presents the step-by-step procedures from a continuous generator model to ARX models along with numerical estimation results. Section VI concludes the paper.


\section{LSE for ARX Model}\label{method}
\begin{figure}[h!]
\centering
\includegraphics[width=0.4\textwidth]{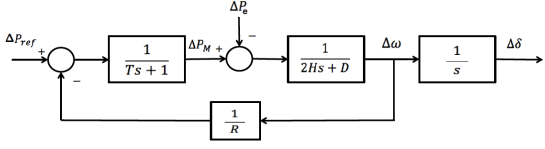}
\caption{A simplified synchronous generator model.} \label{fig:sys}
\end{figure}

%
A block diagram of a simplified synchronous generator model is shown in Figure \ref{fig:sys}, where $\delta$ is the rotor angle (rad), $\omega_{0}$ is the synchronous speed (rad/s), {$H$ is the inertia constant (seconds), $D$ is the damping factor, $R$ is the speed regulation constant (p.u.), $T$ is the time constant of the turbine-governor (seconds),} $P_{m}$ is the mechanical power input (p.u.), and $P_{e}$  is the electrical power (p.u.).

Linearizing the above model, we can develop the Continuous-Time (CT) closed-loop transfer function in Laplace domain, which will be converted into its equivalent Discrete-Time (DT) transfer function in  Z-domain where sampling is considered. The $z$-{transfer} function is then converted into an autoregressive exogenous (ARX) model as follows:

Suppose that the DT transfer function is given by:
\begin{equation}
\frac{y(z)}{u(z)}=\frac{b_{1}z^{-1}+b_{0}z^{-2}}{z+a_{1}z^{-1}+a_{0}z^{-2}}
\end{equation}
where $y$ is the output, and $u$ is the input. Then, its equivalent ARX model can be expressed as:
\begin{equation}\label{ARXX}
y(k)=-a_{1}y(k-1)-a_{0}y(k-2)+b_{1}u(k-1)+b_{0}u(k-2) +e(k)
\end{equation}
where $e$ is the error.

Given time series measurements of the input and output, an overestimated problem can be formulated based on \eqref{ARXX} as follows:
\begin{scriptsize}
\begin{align}\label{matt}
\begin{bmatrix}
y(k) \\ y(k+1) \\ \vdots \\ y(N)
\end{bmatrix}
 = \begin{bmatrix}
{blue}{-}y(k-1) & {-}y(k-2)& u(k-1) & u(k-2)  \\
{-}y(k)   & {-}y(k-1) &u(k) & u(k-1)  \\
\vdots   & \vdots &\vdots &\vdots\\
{-}y(N-1) &{-}y(N-2)  & u(N-1) & u(N-2)
\end{bmatrix}
\begin{bmatrix}
a_1  \\ a_0\\  b_1 \\ b_0
\end{bmatrix}
+e
\end{align}
\end{scriptsize}
In a concise format, \eqref{matt} can be written as $b=Ax+e$. The parameter vector $x$ can be found as $A^T(AA^T)^{-1}b$.

Converting the generator's dynamic model into a difference equation, LSE-based estimation can then be used to estimate the coefficients. After that, the parameters $H$, $D$, $T$, and $R$ can be identified. Before proceeding any further, let's consider how to discretize a CT transfer function given in Laplace domain to find its equivalent $z$-transfer function.
\section{discretization of a ct transfer function}\label{discretization}
Several methods exist for discretizing a CT transfer function given in Laplace Domain. These methods include: 1) Zero-order hold (ZOH) method, where a zero-order hold element is placed at the input of the system to hold the input signal constant during each sampling interval. 2) {Numerical} approximations to time integrals, where Euler's forward, Euler's backward, or Tustin's approximations are used to get substitution {formulas} for the Laplace operator $s$ in terms of $z$ and the sampling period $h$. Although the latter methods are much easier to compute, care should be taken with Euler's forward/ backward approximations because they cannot be used without modifications as discussed in \cite{soderstrom1997least}. For this reason, only ZOH and Tustin's approximation methods are investigated in this paper.
\subsection{Zero-order hold method}
\begin{figure}[h!]
\centering
\includegraphics[width=0.3\textwidth]{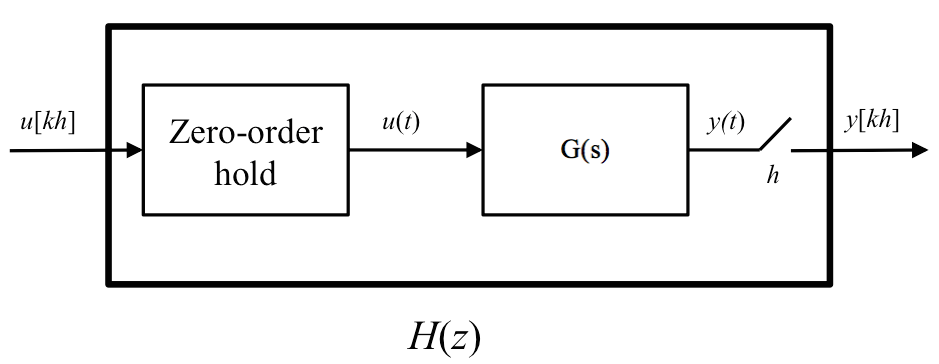}
\caption{{Discretization of a continuous-time system.}} \label{ZOHB}
\end{figure}

For a general linear system, the relationship between the input and output can be determined from the response of the system to a given signal, such as a unit-step input \cite{astrom2011computer}. Let $H(s)$ denote the CT transfer function from the input $u(t)$ to the output $y(t)$. Assuming that the system is preceded by a zero-order hold element and followed by a sampler as shown in Figure \ref{ZOHB}, we wish to find the DT {transfer} function from the sampled input $u(k)$ to the sampled output $y(k)$. This can be achieved by determining the step response of the CT system first. After that, the corresponding $z$-transform of the step response is obtained. Finally, we divide by the $z$-transform of the unit-step function.
\\
This procedure can be formulated as follows:
\begin{equation}\label{eq:pro}
H(z)=\left( \frac{z}{z-1} \right) \mathscr{Z}\left \{ \mathscr{L}^{-1}\left [ \frac{G(s)}{s} \right ] \bigg|_{t=kh} \right \}
\end{equation}
where $\mathscr{Z}$ is the $z$-transform, $\mathscr{L}^{-1}$ is the inverse Laplace, and $h$ is the sampling interval.

\subsection{Tustin's method}
Numerical approximations to the time integral (or derivative) are widely used in digital control applications. [REF 2books]. A DT transfer function in $z$-domain is obtained by using appropriate substitutions to time integrals so that the behaviors of the $z$-transfer function and the CT $s$-transfer function are similar. One of these methods is Tustin's approximation which is also known as the \textit{bilinear transformation}.

In order to {obtain} the {substitution} formula for Tustin's method, consider the simple integrator $ \dot{y}=u$, where $u$ is the input and $y$ is the output. In Laplace domain the above equation can be written as: $\frac{Y(s)}{U(s)}=\frac{1}{s}$.

On the other hand, Tustin's method uses the following approximation to the time integral:
\begin{equation}\label{Tus}
y(k)\approx y(k-1)+\frac{h}{2}\left[u(k)+u(k-1)\right]
\end{equation}
In $z$-domain, the difference equation \eqref{Tus} corresponds to:
\begin{equation}\label{zee}
\frac{y(z)}{u(z)}\approx \frac{h}{2}\frac{z+1}{z-1}
\end{equation}
We finally find the substitution formula:
$
s \leftarrow\frac{2}{h}\frac{z-1}{z+1}$.

\begin{figure}[h!]
\centering
\includegraphics[width=0.3\textwidth]{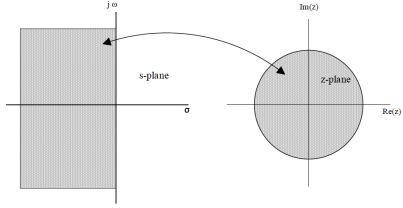}
\caption{Mapping of the stability region for Tustin's approximation.} \label{fig:map}
\end{figure}

As a result, Tustin's approximation maps the left-half of the $s$-domain into the unit circle $|z| < 1$ in $z$-domain as shown in Figure \ref{fig:map}.
\section{Case studies}\label{Cases}
For the case studies, a {benchmark} model based on Figure \ref{fig:sys} is built using MATLAB/Simulink to generate the data where a step response in $P_{e}$ is simulated. The input and output data are collected and then fed into the estimation algorithm to find out the coefficients $a_{i}$ and $b_{i}$. After that, the parameters $H$, $R$, $T$, and $D$ are recovered for several sampling intervals $h$. Different scenarios are simulated and the results are presented for each case. The following parameters are used in the simulation: $H=2.5$, $R=0.05$, and $T=0.5$.

\subsection{Case 1: Power ($\Delta P_{e}$) versus {frequency} ($\Delta \omega$)}
In order to illustrate the procedure and to simplify the computations, the damping factor is ignored (i.e., $D=0$) in this case.
\subsubsection{ZOH method}
Consider the synchronous generator's model shown in Figure \ref{fig:sys}. Without the damping factor, the CT closed-loop transfer function from $\Delta P_{e}$ to $\Delta \omega$ is given by:
\begin{equation}\label{Hc}
H(s)=\frac{\Delta \omega}{\Delta P_{e}}=\frac{Ts + 1}{2HT s^2 +2H s + \frac{1}{R}}
\end{equation}
Since $\frac{2T}{R}>H$, the transfer function represents an underdamped second-order system with two complex conjugate poles given by:
\begin{equation}
s_{1,2}=\frac{-1}{2T}\pm j\ \frac{\sqrt{\frac{2HT}{R}-H^{2}}}{2HT}.
\end{equation}

The transfer function $H(s)/s$ will be fractionated and then inverse Laplace transformation will be applied.

It worth mentioning that even if we include the damping factor, say $D=0.8$ p.u., we still get an {underdamped} response which is generally the case for the synchronous generator under normal operating conditions as discussed in \cite{bergen2000power} Following the procedure given by \eqref{eq:pro}, we obtain the DT transfer function as:
\begin{equation}\label{Zdomain}
H(z)=\frac{y(z)}{u(z)}=\frac{b_{1}z+b_{0}}{z^{2}+a_{1}z+a_{0}}
\end{equation}
Thus, the ARX model can be obtained based on \eqref{Zdomain} as follows:
\begin{equation}\label{ARX2}
y(k+2)=-a_{1}y(k+1)-a_{0}y(k)+b_{1}u(k+1)+b_{0}u(k)
\end{equation}
where $y$ is the output $\Delta \omega$ and $u$ is the input $\Delta P_{e}$,
\begin{align}
\begin{cases}
a_{1}&=-2e^{\frac{-h}{2T}}\cos{(\omega h)}\\
a_{0}&=e^{\frac{-h}{T}}\\
b_{1}&=R\left( 1- e^{\frac{-h}{2T}}\cos{(\omega h)} -ke^{\frac{-h}{2T}} \sin{(\omega h)} \right)\\
b_{0}&=R\left( e^{\frac{-h}{T}} - e^{\frac{-h}{2T}}\cos{(\omega h)} + ke^{\frac{-h}{2T}}\sin{(\omega h)} \right)
\end{cases}
\end{align}
and $\alpha$, $\omega$, and $k$ are given by the following equations:
\begin{align}
\begin{cases}
\alpha&=\frac{HR-T}{2HTR},\\
\omega&=\sqrt{\frac{2T-HR}{4HRT^{2}}},\\
k&=\frac{\alpha}{\omega}.
\end{cases}
\end{align}

If we can estimate the coefficients $a_{1}$, $a_{0}$, $b_{1}$, and $b_{0}$, we can find out the parameters. The ARX model \eqref{ARX2} can be put in the matrix format $b=Ax$ as follows:
\begin{footnotesize}
\begin{align}\label{mat}
\begin{bmatrix}
y(3) \\ y(4) \\ \vdots \\ y(N)
\end{bmatrix}
 = \begin{bmatrix}
{-}y(2) & {-}y(1)& u(2) & u(1)  \\
{-}y(3)   & {-}y(2) &u(3) & u(2)  \\
\vdots   & \vdots &\vdots &\vdots\\
{-}y(N-1) &{-}y(N-2)  & u(N-1) & u(N-2)
\end{bmatrix}
\begin{bmatrix}
a_1  \\ a_0\\  b_1 \\ b_0
\end{bmatrix}
\end{align}
\end{footnotesize}
Using  LSE, the above overestimated problem can be solved by $x=(A^{T}A)^{-1}A^{T}b$. 

After solving the optimization problem and finding the coefficients, the parameters can be found using the following {equations}:
\begin{align}
\begin{cases}
T&=\frac{-h}{\ln{(a_{0})}}\\
R&=\frac{b_{1}+b_{0}}{1+e^{\frac{-h}{T}}+a_{1}}\\
H&=\frac{2T}{R+4RT^{2}\omega^{2}}\\
\omega&=\frac{\cos^{-1}{\left(\frac{-a_{1}e^{\frac{h}{2T}}}{2} \right)}}{h}
\end{cases}
 \end{align}
\subsubsection{Tustin's method}
Obtaining the DT transfer function using Tustin's approximation method is much easier than the ZOH method {because} we do not need to worry about the poles or the system's type of response. Substitute the {Laplace} operator $s$ in \eqref{Hc}, we get:
\begin{equation}\label{Ztustin's}
H(z)=\frac{y(z)}{u(z)}=\frac{b_{2}z^{2}+b_{1}z+b_{0}}{z^{2}+a_{1}z+a_{0}},
\end{equation}
where the coefficients are given by:
\begin{align}
\begin{cases}
a_{1}&=\frac{2-4HRTk^{2}}{\alpha} \\
a_{0}&=\frac{2HRTk^{2}-2HRk+1}{\alpha}\\
b_{2}&=\frac{R(1+Tk)}{\alpha}\\
b_{1}&=\frac{2R}{\alpha}\\
b_{0}&=\frac{R(1-Tk)}{\alpha}
\end{cases}
\end{align}
and $\alpha=2HRTk^2+2HRk+1$, {$k=\frac{2}{h}$}.
After that, the ARX model is developed and the optimization problem is solved as discussed before. Once the coefficients are found, the parameters can be easily found. Expressions of $b_1$ and $b_2$ are used to find $T$  as follows:
\begin{align}\label{Tee}
\begin{cases}
T&=\frac{2B-1}{k},\\
B&=\frac{b_{2}}{b_{1}}
\end{cases}
\end{align}
then, using the expression of $a_1$ we can find $R$ in terms of $H$ as:
\begin{align}\label{RinH}
R&=\frac{2-a_{1}}{H\left(2a_{1}Tk^{2}+2a_{1}k+4Tk^{2}\right)}
\end{align}
Finally, we can {substitute} \eqref{Tee} and \eqref{RinH} in the expression of $b_2$ to find $H$ as follows:
\begin{align}
H=\frac{1+Tk-\frac{b_{2}}{R}}{2b_{2}k(Tk+1)}
\end{align}
\subsubsection{Numerical results}
\begin{figure}[h!]
\centering
\includegraphics[width=0.4\textwidth]{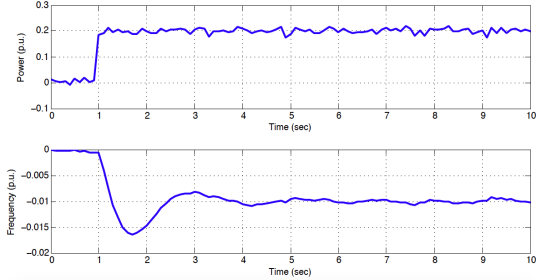}
\caption{Change in power $\Delta P_{e}$ and corresponding change in frequency $\Delta \omega$.}
\label{inout}
\end{figure}

The benchmark model is simulated using MATLAB/Simulink for various input signals and the parameters are estimated. Noise with 0.0001 variance and zero mean is added to the input signal. A step input equals 0.2 p.u. is applied at $t=1$ second and the p.u. change in the input power ($\Delta P_{e}$) and output {frequency} ($\Delta \omega$) are shown in Figure \ref{inout} when $h=0.1$ second.
\\
The estimated coefficients using Matlab Optimization toolbox CVX \cite{cvx} when $h=0.1$ are shown in Table \ref{tab:coeffi1}. And the recovered parameters for different sampling intervals are shown in Table \ref{tab:rec1}.
\begin{table}[h]
\caption {estimated coefficients using CVX} \label{tab:coeffi1}
\centering
\begin{tabular}{|c|c|c|}
ine
Coefficient & ZOH method         & Tustin's method    \\ ine
$b_{2}$          & x                  & 9.8214$\times 10^{-3}$  \\ ine
$b_{1}$          & 19.7333$\times 10^{-3}$  & 1.7857$\times 10^{-3}$  \\ ine
$b_{0}$         & -16.1380$\times 10^{-3}$ & -8.0357$\times 10^{-3}$  \\ ine
$a_{1}$         & -1.7467 & -1.7500 \\ ine
$a_{0}$          & 818.5742$\times 10^{-3}$  & 821.4286$\times 10^{-3}$  \\ ine
\end{tabular}
\end{table}
\begin{table}[h]
\caption {Recovered parameters for different $h$} \label{tab:rec1}
\centering
\begin{tabular}{|c|c|c|l|l|l|l|}
ine
  & \multicolumn{3}{c|}{ZOH method} & \multicolumn{3}{c|}{Tustin's method} \\ ine
  & h=0.1    & h=0.01   & h=0.001   & h=0.1      & h=0.01     & h=0.001    \\ ine
T & 0.499    & 0.499    & 0.500     & 0.500      & 0.500      & 0.500      \\ ine
R & 0.050    & 0.050    & 0.050     & {0.0500}      & {0.0499}      & {0.0499}      \\ ine
H & 2.506    & 2.500    & 2.499     & {2.500}     & {2.499}     & {2.499}     \\ ine
\end{tabular}
\end{table}

It should be mentioned that in order for Tustin's method to work, the CT model in Laplace domain cannot be simulated directly, otherwise, the estimated coefficients are almost equal to the coefficients of ZOH method with $b_{2}=0$. Discrete transfer function blocks shall be used where each $s$ is replaced by $\frac{2}{h}\frac{z-1}{z+1}$; this way correct values of the parameters are obtained.
\subsection{Case 2: Power ($\Delta P_{e}$) versus rotor angle ($\Delta \delta$)}
\begin{figure}[h!]
\centering
\includegraphics[width=0.4\textwidth]{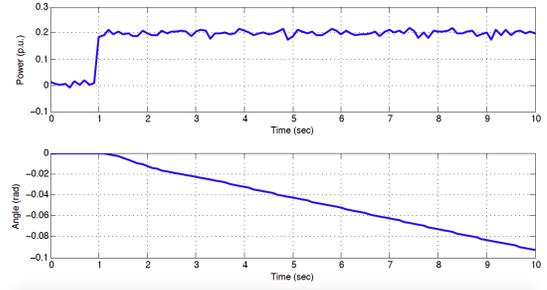}
\caption{Change in power $\Delta P_{e}$ and corresponding change in angle $\Delta \delta$.}
\label{fig:pangle}
\end{figure}

In this case, the change in the rotor angle $\Delta \delta$ is considered as the output while the change in the electrical power $\Delta P_{e}$ is the input. The damping factor is again {ignored} in this case. The same input of case 1 is applied at $t=1$ second and the p.u. change in both power and angle are shown in Figure \ref{fig:pangle} for $h=0.1$ second.
Since the procedure is explained in the first case, it will not be covered here.
\subsubsection{ZOH method}
Following the aforementioned procedure, the DT transfer function has the form:
\begin{equation}\label{Zforfelta}
H(z)=\frac{y(z)}{u(z)}=\frac{b_{2}z^{2}+b_{1}z+b_{0}}{z^{3}+a_{2}z^{2}+a_{1}z+a_{0}}
\end{equation}
where:
\begin{align}
\begin{cases}
a_{2}&=-2e^{\frac{-h}{2T}}\cos{(\omega h)}-1\\
a_{1}&=e^{\frac{-h}{T}}+2e^{\frac{-h}{2T}}\cos{(\omega h)}\\
a_{0}&=-e^{\frac{-h}{T}}\\
b_{2}&=R\left[a-ae^{\frac{-h}{2T}}\cos{(\omega h)}-be^{\frac{-h}{2T}}\sin{(\omega h)}+h\right]\\
b_{1}&=R[-a+ae^{\frac{-h}{T}}-2he^{\frac{-h}{2T}}\cos{(\omega h)}+2be^{\frac{-h}{2T}}\sin{(\omega h)}]\\
b_{0}&=R\left[(h-a)e^{\frac{-h}{T}}+ae^{\frac{-h}{2T}}\cos{(\omega h)}-be^{\frac{-h}{2T}}\sin{(\omega h)}\right]\\
\end{cases}
\end{align}
and $a=T-2HR$, $b=\frac{\alpha}{\omega}$, $\alpha=\frac{3T-2HR}{2T}$, and $\omega=\sqrt{\frac{2T-HR}{4HRT^{2}}}$.

After estimating the coefficients, the parameters can be identified. The recovered parameters for different sampling intervals are listed in Table \ref{tab:angle}.
\subsubsection{{Tustin's method}} 
\begin{figure}[h!]
\centering
\includegraphics[width=0.4\textwidth]{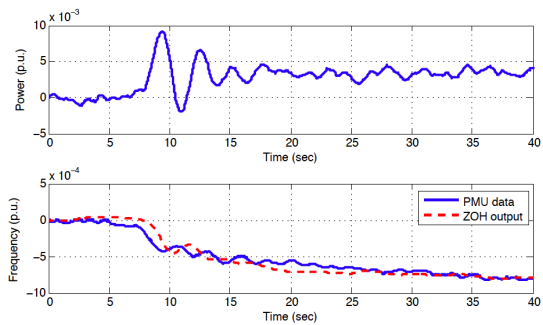}
\caption{Real-world PMU measurements versus ZOH estimation.}
\label{fig:rea}
\end{figure}

{The DT transfer function is in the form:
\begin{equation}\label{Zfordelta}
H(z)=\frac{y(z)}{u(z)}=\frac{b_{3}z^{3}+b_{2}z^{2}+b_{1}z+b_{0}}{z^{3}+a_{2}z^{2}+a_{1}z+a_{0}}
\end{equation}
where:
\begin{align}
\begin{cases}
a_{2}&=\frac{-6HRTk^3-2HRk^2+k}{\alpha}\\
a_{1}&=\frac{6HRTk^3-2HRk^2-k}{\alpha}\\
a_{0}&=\frac{-2HRTk^3+2HRk^2-k}{\alpha}\\
b_{3}&=\frac{R(1+Tk)}{\alpha}\\
b_{2}&=\frac{R(3+Tk)}{\alpha}\\
b_{1}&=\frac{R(3-Tk)}{\alpha}\\
b_{0}&=\frac{R(1-Tk)}{\alpha}\\
\end{cases}
\end{align}
and $k=\frac{2}{h}$, $\alpha=2HRTk^3+2HRk^2+k$. The recovered parameters after estimating the coefficients are shown in Table \ref{tab:angle}.}
\begin{table}[h]
\centering
\caption{Recovered parameters using ZOH and Tustin's methods}
\label{tab:angle}
\begin{tabular}{|c|c|c|c|c|c|c|}
ine
  & \multicolumn{3}{c|}{ZOH method} & \multicolumn{3}{c|}{{Tustin's method}} \\ ine
  & h=0.1    & h=0.01   & h=0.001   & {h=0.1}     & {h=0.01}     & {h=0.001}     \\ ine
T & 0.499    & 0.499    & 0.499     & {0.499}     & {0.500}      & {0.500}       \\ ine
R & 0.050    & 0.050    & 0.0499    & {0.050}     & {0.0499}     & {0.0498}      \\ ine
H & 2.501    & 2.500    & 2.499     & {2.499}     & {2.500}      & {2.500}       \\ ine
\end{tabular}
\end{table}
{
\subsection{Case 3: Real-world PMU data}
In this case both ZOH and Tustin's methods, are used to estimate the parameters from a recorded real-world PMU data of an anonymous busbar of MISO system. The data is recorded for about 40 seconds when a major disturbance occurs in the system. Similar to case 1, the change in power and frequency are used for estimating the parameters. The estimated parameters for both methods are listed in Table \ref{table:estimatedpmu} which clearly shows that Tustin's method fails to provide correct estimation of the parameters. In order to validate the results of ZOH method, the model is built in MATLAB/Simulink and event playback is used to inject the change in power as input. The change in frequncy is then compared to the PMU measurement as shown in Figure \ref{fig:rea} which demonstrates a great degree of match.}
\begin{table}[h]
\centering
\caption{Recovered parameters for real-world PMU data}
\label{table:estimatedpmu}
\begin{tabular}{|c|c|c|c|}
ine
                & T      & R        & H       \\ ine
ZOH method      & 0.4534 & 0.2320   & 13.8945 \\ ine
Tustin's method & 0.001  & 0.000667 & 0.0639  \\ ine
\end{tabular}
\end{table}

\section{conclusion}
\label{conclusion}
In this paper, linear LSE-based generator model parameter estimation via PMU data has been presented in detail. The continuous time Laplace model is first converted to a discrete ARX model using ZOH method or Tustin's method. The coefficients of the ARX model will then be identified using linear LSE. The generator model parameters are then recovered by the ARX model coefficients. The proposed approaches are illustrated in numerical case studies to demonstrate their effectiveness in identifying generator parameters. It is found that for real-world data, ZOH method is robust while Tustin's method is very sensitive to the estimation model.
\bibliographystyle{IEEEtran}
\bibliography{IEEEabrv,y}

\end{document}